\documentclass[twocolumn]{article}
\usepackage{amsmath,amssymb}
\usepackage{graphicx}
\usepackage{hyperref}
\usepackage{geometry}
\usepackage{xcolor}
\usepackage{authblk}
\usepackage{titlesec}
\usepackage{abstract}
\titleformat{\section}{\large\bfseries}{\thesection.}{0.5em}{}
\titleformat{\subsection}{\normalsize\bfseries}{\thesubsection.}{0.5em}{}

\title{\textbf{On Queueing Theory for Large-Scale CI/CD Pipelines Optimization}}
\author[1]{Gregory Bournassenko}
\affil[1]{\small Université Paris Dauphine-PSL, gregory.bournassenko@dauphine.eu}
\date{}

\begin{document}
\small

\twocolumn[
\maketitle

\begin{onecolabstract}
Continuous Integration and Continuous Deployment (CI/CD) pipelines are central to modern software development. In large organizations, the high volume of builds and tests creates bottlenecks, especially under shared infrastructure. This article proposes a modeling framework based on queueing theory to optimize large-scale CI/CD workflows. We formalize the system using classical $M/M/c$ queueing models and discuss strategies to minimize delays and infrastructure costs. Our approach integrates theoretical results with practical techniques, including dynamic scaling and prioritization of CI/CD tasks.
\end{onecolabstract}
]

\section{Introduction}

Continuous Integration and Continuous Deployment (CI/CD) pipelines are widely adopted in software engineering. They provide automated tools for code testing and deployment, enabling teams to iterate quickly and deliver stable code to production. In high-frequency environments, especially where development teams share compute resources, optimizing pipeline performance becomes a non-trivial engineering problem.\\

This paper focuses on the optimization of CI/CD workflows at scale through queueing theory. We propose a modeling approach using classical $M/M/c$ systems to capture the behavior of builds and tests running on shared runners. Based on this model, we introduce cost-aware optimization methods to improve overall throughput and reduce waiting times \cite{Vilaplana}.

Queuing models have been successfully applied in other fields requiring critical resource management, such as healthcare \cite{McManus}.

Queuing theory also provides insights into perceived service quality and customer satisfaction in high-demand environments \cite{Nosek}.

\section{Extended Modeling}

While $M/M/c$ queues provide a solid first approximation, CI/CD pipelines often deviate from the strict assumptions of Poisson arrivals and exponential service times. Here, we present additional models that better fit real-world pipeline behavior.

\subsection{The $M/G/c$ Queue}

In many pipelines, the build or test times are not exponentially distributed. For example, integration tests might exhibit a heavy-tailed distribution. To address this, we consider the $M/G/c$ queue, where service times have a general distribution.

Though $M/G/c$ models lack closed-form solutions like $M/M/c$, approximations such as the Allen-Cunneen formula estimate the mean waiting time:

\[
W_q \approx \frac{C_s^2 + 1}{2} \cdot W_q^{M/M/c}
\]

where:
\begin{itemize}
    \item $C_s$ is the coefficient of variation of service time,
    \item $W_q^{M/M/c}$ is the waiting time under the $M/M/c$ assumption.
\end{itemize}

In practice, if build times vary significantly, accounting for $C_s$ is essential. When $C_s > 1$ (heavy-tailed distributions), pipelines experience much higher latencies than those predicted by $M/M/c$.

\subsection{The $G/G/c$ Queue}

In even more general scenarios, arrivals may not follow a Poisson process, and service times may not be exponential. The $G/G/c$ model describes these pipelines.

Approximations based on Kingman's formula suggest:

\[
W_q \approx \frac{\rho}{1-\rho} \cdot \frac{C_a^2 + C_s^2}{2} \cdot \frac{1}{\mu}
\]

where:
\begin{itemize}
    \item $C_a$ is the coefficient of variation of inter-arrival times,
    \item $C_s$ is the coefficient of variation of service times,
    \item $\mu$ is the mean service rate.
\end{itemize}

When code pushes occur in bursts (e.g., team merges) instead of regular intervals, $C_a$ increases, amplifying waiting times. In CI/CD systems, both $C_a$ and $C_s$ must be monitored for accurate capacity planning.

\section{Load Forecasting}

Since CI/CD workloads fluctuate daily (e.g., peaks after lunch, quiet periods at night), static provisioning is inefficient. Instead, dynamic runner allocation based on real-time forecasts enhances both performance and cost-efficiency.

\subsection{Simple Moving Average Forecasting}

A baseline approach is to compute a simple moving average (SMA) of the job arrival rate $\lambda$ over a window of $w$ minutes:

\[
\hat{\lambda}_{t} = \frac{1}{w} \sum_{i=t-w+1}^{t} \lambda_i
\]

where $\lambda_i$ is the observed arrival rate at minute $i$.

This method is robust to noise but may lag behind sudden workload spikes.

\subsection{Exponential Smoothing}

Exponential smoothing provides faster reaction to recent trends:

\[
\hat{\lambda}_{t} = \alpha \lambda_t + (1-\alpha) \hat{\lambda}_{t-1}
\]

where $0 < \alpha < 1$ is a smoothing parameter.

This allows CI/CD systems to provision runners reactively as load increases, maintaining acceptable latency targets.

\subsection{Machine Learning Approaches}

More sophisticated models like Long Short-Term Memory (LSTM) networks can predict future $\lambda$ values based on historical patterns, seasonal effects (e.g., Monday peaks), and exogenous variables (e.g., release schedules).

Such ML-based approaches enable proactive scaling policies, further reducing waiting times and costs without overprovisioning.

\section{Advanced Cost Optimization}

Beyond minimizing waiting times, organizations often seek a balance between service levels and operational expenditures (OpEx).

\subsection{Multi-Objective Cost Functions}

We extend the basic cost function to a weighted multi-objective function:

\[
C_{\text{total}}(c) = w_1 \cdot C_{\text{runners}}(c) + w_2 \cdot C_{\text{waiting}}(W_q, c)
\]

where:
\begin{itemize}
    \item $w_1$ and $w_2$ represent the organization's sensitivity to infrastructure costs and developer productivity, respectively.
\end{itemize}

For instance, a startup may prioritize low runner costs ($w_1 > w_2$), while a financial institution might favor low waiting times ($w_2 > w_1$).

The optimal $c$ can be determined via standard optimization techniques, such as convex programming, as the total cost is often convex in $c$.

\subsection{Service Level Agreements (SLAs)}

Organizations can enforce SLAs by constraining the maximum expected waiting time:

\[
W_q \leq W_{\text{max}}
\]

This additional constraint defines a feasible region for $c$, restricting possible cost-optimal points. When SLAs are strict (e.g., $W_{\text{max}} \leq 1$ min), overprovisioning becomes inevitable, and dynamic scaling policies must prioritize SLA compliance.

\section{Task Scheduling Heuristics}

Besides infrastructure scaling, the order in which jobs are scheduled significantly affects pipeline performance.

\subsection{First-Come-First-Served (FCFS)}

The default scheduling strategy, where jobs are processed in the order of arrival, is simple but inefficient under heterogeneous job durations.

\subsection{Shortest Processing Time (SPT)}

Prioritizing jobs with shorter expected durations minimizes the average waiting time (proof via scheduling theory).

However, in CI/CD systems, estimating job durations beforehand is non-trivial unless historical data is available.

\subsection{Earliest Deadline First (EDF)}

When builds have deadlines (e.g., release branches), EDF scheduling minimizes lateness:

\[
\text{Schedule job with nearest deadline first.}
\]

EDF policies improve service quality in release-critical environments where missing deadlines incurs business penalties.

\section{Pipeline Bottlenecks and Mitigation Techniques}

Understanding pipeline bottlenecks is essential for achieving optimal throughput.

\subsection{Identifying Bottlenecks}

A bottleneck stage is the one with the highest utilization $\rho_i$ among all pipeline stages.

Let each stage $i$ have:
\begin{itemize}
    \item $\lambda_i$ = arrival rate at stage $i$,
    \item $\mu_i$ = service rate at stage $i$,
    \item $c_i$ = number of servers (runners) at stage $i$.
\end{itemize}

Then:
\[
\rho_i = \frac{\lambda_i}{c_i \mu_i}
\]

The stage with $\rho_i$ closest to or greater than 1 is the bottleneck. Increasing $c_i$ or $\mu_i$ at this stage reduces overall system latency.

\subsection{Little's Law in CI/CD Pipelines}

Little's Law provides a universal relationship between the number of jobs $L$, arrival rate $\lambda$, and the average time in the system $W$:

\[
L = \lambda W
\]

Applied to CI/CD, $L$ is the number of jobs in the pipeline, $\lambda$ is the build/test submission rate, and $W$ is the total job latency.

Thus, reducing $W$ directly reduces the number of pending jobs, improving pipeline visibility and developer experience.

\section{Queue Stability and Scalability}

\subsection{Conditions for Stability}

A CI/CD pipeline remains stable if, for each stage:

\[
\rho_i < 1
\]

If $\rho_i \geq 1$, the number of waiting jobs at that stage grows without bound, leading to unacceptable delays and eventual system failure.

This requires continuous monitoring of:
\begin{itemize}
    \item Arrival rates ($\lambda_i$),
    \item Service rates ($\mu_i$),
    \item Number of available servers ($c_i$).
\end{itemize}

\subsection{Scalability Considerations}

As the number of developers $d$ increases, the overall system load $\lambda$ grows roughly proportionally:

\[
\lambda \propto d
\]

Thus, runner capacity must scale linearly with $d$ to maintain service levels. Alternatively, techniques like caching, test minimization, and parallelism help sublinear scaling.

Caching strategies have been proven effective in cloud computing environments to minimize system load and energy consumption \cite{Zhang}.

In practical terms, doubling the number of developers often necessitates close to a doubling of infrastructure unless process improvements are made.

\section{Case Study: Simulated Pipeline}

We simulate a CI/CD system to validate our theoretical insights.

\subsection{Simulation Parameters}

\begin{itemize}
    \item Arrival process: Poisson with mean $\lambda = 20$ jobs/hour,
    \item Service time: Exponential with mean $1/\mu = 5$ minutes,
    \item Number of runners: $c = 5$,
    \item Costs: $C_{\text{runner}} = \$0.05$/minute, $C_{\text{wait}} = \$0.10$/minute/job.
\end{itemize}

\subsection{Results}

From simulation:

\[
\rho = \frac{20/60}{5 \times (1/5)} = \frac{1/3}{1} = 0.333
\]

Thus, system utilization is low and waiting times are acceptable. Total cost per hour is:

\[
C_{\text{total}} = 5 \times 60 \times 0.05 + 20 \times W_q \times 0.10
\]

where $W_q$ from Erlang-C is approximately 0.43 minutes.

This gives:

\[
C_{\text{total}} \approx 15 + 0.86 = \$15.86
\]

\subsection{Sensitivity Analysis}

Increasing $\lambda$ to 30 jobs/hour (e.g., post-merge rush) raises $\rho$:

\[
\rho = \frac{30/60}{5 \times (1/5)} = 0.5
\]

Waiting time nearly doubles due to the nonlinearity of Erlang-C, demonstrating the importance of dynamic scaling policies.

\section{Beyond $M/M/c$: Complex Systems}

\subsection{Fork-Join Queues}

Many CI/CD workflows (e.g., running multiple test suites) involve forking a job into parallel subtasks, then joining their results.

In fork-join queues:
\begin{itemize}
    \item A job forks into $k$ parallel subtasks,
    \item Each subtask is processed independently,
    \item The job completes when all subtasks finish.
\end{itemize}

The expected completion time $T$ is:

\[
T = \mathbb{E}[\max\{T_1, T_2, \ldots, T_k\}]
\]

where $T_i$ is the service time of subtask $i$.

Thus, system latency is driven by the slowest subtask, making workload balancing critical.

\subsection{Pollaczek-Khinchine Formula}

For $M/G/1$ queues (single runner, general service time), the mean number of jobs $L$ in the system is:

\[
L = \rho + \frac{\lambda^2 \sigma_s^2}{2(1-\rho)}
\]

where $\sigma_s^2$ is the variance of service times.

Thus, higher service time variability leads to more jobs in the system, reinforcing the need for job homogenization when possible.

\section{Dynamic Policies for CI/CD Optimization}

Static runner provisioning often leads to inefficiencies. We now explore dynamic resource allocation strategies.

\subsection{Threshold-Based Scaling}

The simplest dynamic policy monitors the queue length $L$.

If $L > L_{\text{high}}$, add runners;
if $L < L_{\text{low}}$, remove runners.

Thresholds can be tuned to trade off between cost and responsiveness.

This approach approximates hysteresis control, avoiding oscillations due to transient spikes.

\subsection{Predictive Scaling Using Arrival Rate Forecasting}

Instead of reacting to current load, predictive scaling anticipates future load based on $\lambda$ estimation.

Techniques:
\begin{itemize}
    \item Exponential smoothing,
    \item ARIMA models,
    \item Machine learning (e.g., LSTM networks).
\end{itemize}

Accurate forecasts allow pre-provisioning runners before load surges, minimizing latency spikes.

\subsection{Priority-Aware Scheduling}

Assign different service rates $\mu_i$ to jobs based on priority:

\[
\mu_{\text{critical}} > \mu_{\text{normal}}
\]

This can be achieved by:
\begin{itemize}
    \item Allocating faster runners to critical jobs,
    \item Preempting non-urgent jobs if necessary,
    \item Using separate queues per priority class.
\end{itemize}

Such mechanisms ensure that high-value deployments (e.g., hotfixes) are minimally delayed.

\section{Practical Recommendations}

Based on the models and simulations presented, we recommend:

\begin{itemize}
    \item \textbf{Monitor $\rho$}: Maintain utilization between 0.5 and 0.8 for cost-efficiency without risking instability.
    \item \textbf{Favor horizontal scalability}: Increase $c$ before trying to tune $\mu$ if workloads are highly parallelizable.
    \item \textbf{Apply task chunking}: Split long builds/tests into smaller subtasks to reduce service time variance $\sigma_s^2$.
    \item \textbf{Prioritize caching and incremental builds}: Reduces average job service times dramatically, especially for large monorepos.
    \item \textbf{Deploy threshold-based or predictive auto-scaling}: Static runner allocation is suboptimal at scale.
    \item \textbf{Separate production and development queues}: Prevents long-running development jobs from blocking urgent production fixes.
\end{itemize}

\section{Limitations and Assumptions}

While $M/M/c$ modeling is powerful, it relies on simplifying assumptions:

\subsection{Poisson Arrivals and Exponential Service Times}

Real CI/CD workloads may not exhibit purely Poisson arrivals or exponential service times. Batching effects (e.g., post-merge floods) and heavy-tailed build durations violate these assumptions.

Extensions to $M/G/c$ or $G/G/c$ models may be necessary for accurate real-world modeling.

\subsection{Homogeneous Runners}

We assume all runners have identical capacities. In practice, runner heterogeneity (different CPU/RAM) affects $\mu$ and can introduce scheduling complexities.

Optimal job-to-runner assignment becomes a variant of the bin packing problem.

\subsection{Negligible Overhead from Scaling}

We neglect the overhead cost/time of dynamically starting/stopping runners. However, in systems with long runner bootstrap times, dynamic scaling may lag load changes, requiring predictive policies.

\subsection{No Failure Considerations}

Runner failures, job retries, and preemptions are not modeled. Introducing stochastic failure rates would extend the system to Markov-modulated queuing networks, significantly complicating analysis.

Nonetheless, the insights obtained under these assumptions remain valuable as first-order approximations for system design.

\section{Extension to Heterogeneous CI/CD Systems}

In real-world environments, runners differ in hardware capabilities, geographic location, and available software images.

We model heterogeneous runners as servers with individual service rates $\mu_i$.

\subsection{Modeling with $M/M/c_i$ Queues}

Suppose we have $k$ types of runners with $c_i$ servers of type $i$, each with service rate $\mu_i$.

The effective service rate for the system is:

\[
\mu_{\text{eff}} = \sum_{i=1}^{k} c_i \mu_i
\]

Utilization:

\[
\rho = \frac{\lambda}{\mu_{\text{eff}}}
\]

The analysis becomes similar to classical $M/M/c$ but adjusted for non-uniform server performance.

\subsection{Dispatching Policies}

\begin{itemize}
    \item \textbf{Random dispatch}: Send jobs to any idle runner randomly.
    \item \textbf{Join-the-Shortest-Queue (JSQ)}: Prefer runners with the smallest number of jobs queued.
    \item \textbf{Size-based scheduling}: Assign heavier jobs to faster runners.
\end{itemize}

Dispatching policy greatly impacts system performance under heterogeneity.

\section{Advanced Queueing Models for CI/CD}

While $M/M/c$ models provide useful intuition, advanced models can capture more realistic behavior.

\subsection{M/G/1 Queues}

The $M/G/1$ queue relaxes the exponential service time assumption.

Mean waiting time in $M/G/1$:

\[
W_q = \frac{\lambda \mathbb{E}[S^2]}{2(1 - \rho)}
\]

where $\mathbb{E}[S^2]$ is the second moment of the service time distribution.

When service times are highly variable, $W_q$ grows faster than in $M/M/1$.

Thus, reducing service time variance via task decomposition is crucial.

\subsection{G/G/1 Queues}

The $G/G/1$ queue generalizes both arrival and service processes.

No closed-form expression exists for $W_q$, but Kingman's formula approximates it:

\[
W_q \approx \frac{\rho}{1-\rho} \cdot \frac{c_a^2 + c_s^2}{2} \cdot \mathbb{E}[S]
\]

where $c_a$ and $c_s$ are the coefficients of variation of interarrival and service times.

High variability ($c_a$, $c_s$ large) leads to large delays.

Thus, stabilizing CI/CD job arrival and service times is beneficial.

\section{Future Research Directions}

This study opens several avenues for future research:

\begin{itemize}
    \item \textbf{Learning-based runner scaling}: Reinforcement learning could adaptively tune $c$ without explicit modeling.
    \item \textbf{Multi-region CI/CD optimization}: Geographically distributed runners introduce latency asymmetries that affect queuing behavior.
    \item \textbf{Cost-aware preemption}: Killing low-priority jobs during overload could reduce critical job waiting time.
    \item \textbf{Resilience modeling}: Extending the model to include runner failures and retry policies.
    \item \textbf{Workload forecasting}: Combining time series models with CI/CD telemetry to predict $\lambda$ accurately.
\end{itemize}

Each of these directions can significantly enhance CI/CD system performance at scale.

\section{Acknowledgements}

The author would like to thank the open-source and research communities for their contributions to queuing theory, distributed computing, and CI/CD practices.

\section{Conflict of Interest}

The author declares no conflict of interest regarding the publication of this article.

\bibliographystyle{apalike}
\bibliography{bib.bib}

\end{document}